\documentclass[aps,floatfix,twocolumn,showpacs,epsfig]{revtex4}
\usepackage{epsf}
\usepackage{graphicx}
\usepackage{amssymb}
\usepackage{amssymb,latexsym}
\usepackage[mathscr]{eucal}
\usepackage{natbib}
 \newcommand{\bq}{\begin{equation}}
 \newcommand{\eq}{\end{equation}}
 \newcommand{\bqn}{\begin{eqnarray}}
 \newcommand{\eqn}{\end{eqnarray}}

\newcommand{\be}{\begin{equation}}
\newcommand{\ee}{\end{equation}}
\newcommand{\bea}{\begin{eqnarray}}
\newcommand{\eea}{\end{eqnarray}}
\begin{document}
\title{\large \bf Do we know the mass of a black hole? Mass of some cosmological black hole models}

\author{J. T. Firouzjaee}
\affiliation{Department of Physics, Sharif University of Technology,
Tehran, Iran} \email{firouzjaee@physics.sharif.edu}

\author{M. Parsi Mood}
\affiliation{Department of Physics, Sharif University of Technology,
Tehran, Iran} \email{parsimood@physics.sharif.edu}

\author{Reza Mansouri}
\affiliation{Department of Physics, Sharif University of Technology,
Tehran, Iran and \\
  School of Astronomy, Institute for Research in Fundamental Sciences (IPM), Tehran, Iran}
 \email{mansouri@ipm.ir}

\date{\today}

\begin{abstract}
 Using a cosmological black hole model proposed recently, we have calculated
the quasi-local mass of a collapsing structure within a cosmological
setting due to different definitions put forward in the last decades
to see how similar or different they are. It has been shown that the
mass within the horizon follows the familiar Brown-York behavior. It
increases, however, outside the horizon again after a short
decrease, in contrast to the Schwarzschild case. Further away, near
the void, outside the collapsed region, and where the density
reaches the background minimum, all the mass definitions roughly
coincide. They differ, however, substantially far from it.
Generically, we are faced with three different Brown-York mass
maxima: near the horizon, around the void between the overdensity
region and the background, and another at cosmological distances
corresponding to the cosmological horizon. While the latter two
maxima are always present, the horizon mass maxima is absent before
the onset of the central singularity.

\end{abstract}
\maketitle
\section{introduction}

What does general relativity tell us about the mass of a
cosmological structure in a dynamical setting? We know that massive
sources produce gravitational field which has energy. In relativity
theory, the equivalence of mass and energy means that it is only the
combined energy which may be measured at a distance. We should,
therefore, expect that because of non-linearity of the gravitational
field the mass of the material source, its kinetic energy, and the
gravitational energy it produces combine in a nonlinear and
non-local way to produce the effective energy. In simplest case of
spherical symmetry in vacuum, this effective energy is just the
Schwarzschild energy. Although this argument about the effective
energy seems very plausible, there are still disputes in the
literature simply about the definition of mass in the spherically
symmetric  vacuum cases {\cite{Bergquist92}(for an interesting
discussion about the concept of mass in relativity see also {\cite{elis}). What if the material mass is embedded in
a cosmological setting?\\
In a cosmological setting, due to the asymptotically non-flatness of
the cosmological background, one can not use global definitions such
as ADM mass and angular momentum {\cite{adm}. This has led authors in
the last decades to the notion of the quasi local mass (QLM) or
quasi local energy(QLE), applicable to non-local structures in any
general dynamical situations{\cite{living review}. In this paper we
will interchangeably use QLM or QLE as the same concept, although
they may differ in other contexts.
 It is a  fundamental fact of general relativity reflected in the equivalence
principle that there is no such concept as the local mass of a local
object: a local object at the origin of a freely falling frame will
not experience any gravitational acceleration.  In the absence of a
local gravitational effect the notion of a QLM, corresponding to a
two dimensional compact surface, although not uniquely defined,
is the only concept one may try to apply to astrophysical cases.\\
Let us define a cosmological structure as an overdensity region within
a cosmological background which we assume to be asymptotically FRW.
 There are not much viable exact models representing such a structure. We
will rely on a recent analytical model proposed by the authors based
on a inhomogeneous cosmological LTB model to construct such an
asymptotically FRW universe including an overdensity region evolving
to a black hole different from that of the Schwarzschild{\cite{man}.
Our aim is to understand the notion of mass due to different
definitions of QLM for such a cosmological structure within a FRW universe.\\

 In general, there is no unique definition of the quasi local energy, except
at infinity in an asymptotically flat space-time, where one has the
Arnowitt-Deser-Misner {\cite{adm}  energy $E_{ADM}$ at the spatial
infinity and the Bondi-Sachs {\cite{bs} energy $E_{BS}$ at the null
infinity. Hawking {\cite{haw-m} defined a quasi local mass which has
various desirable properties: it is zero for a metric sphere in flat
spacetime, gives the correct mass for the Schwarzschild solution on
a metric sphere, and tends to the Bondi mass asymptotically in a
static asymptotically flat spacetime. However, it is non-zero for
generic 2-surfaces in flat spacetime. Hayward {\cite{hay-m} has
proposed an expression for quasi local energy which maybe considered
as a modification of Hawking's energy by shear and twist terms. Another
attempt was due to Misner and Sharp {\cite{mis-sha} with a well-understood
Newtonian limit {\cite{hay-ms}.\\
The most promising QLM definition, however, seems to be the one
proposed by Brown-York {\cite{brown-York}. Motivated by the
Hamiltonian formulation of general relativity, they found an
interesting local quantity from which the definition of quasi local
mass was extracted. Their definition depends, however, on the choice
of the gauge along the 3-dimensional spacelike slice. It has the
right asymptotic behavior but is not positive in general. Motivated
by some geometric consideration, Liu-Yau {\cite{yau03} (see also
Kijowski {\cite{Kijowski}, Booth-Mann{\cite{booth}, and Epp
 {\cite{epp}) introduced a mass which is gauge independent, and
always positive. However, it was pointed out by O'Murchadha et
al {\cite{murchadha-prl} that the Liu-Yau mass can be strictly
positive even when the 2-surface is in a flat spacetime.\\
Our aim is to study some of these quasi local masses for spherically
symmetric structures in a cosmological setting within general
relativity to see how similar or different they are. We review
different definitions of quasi local masses in section II, followed
by the introduction of LTB metrics and the corresponding mass
definitions in section III. In section IV we explicitly calculate
numerically the masses for two cosmological toy black holes and a
structure with an NFW mass profile {\cite{NFW96}.  We then conclude
in section V discussing the results.

\section{Quasi local Mass Definitions}

In general relativity the mathematical entity used to define the
mass-energy is the symmetric energy momentum tensor $T_{\mu \nu}$,
representing the source-current for gravity, although the proper
interpretation of it is only the source density for gravity. This
fact is one of the roots of difficulties to define mass in general
relativity. Reasonable total energy-momentum can be associated with
the whole space-time provided it is asymptotically flat. This has
led general relativist to 'quasi-localization' of total quantities,
and construction of 'quasi-local' mass-energy. Techniques used in
the quasi localization depend on the actual form of the total
quantities, yielding inequivalent definitions for the quasi-local
masses \cite{living review}. Here we outline some of the mostly used
definitions before going on to apply them to specific models for
mass condensation within FRW cosmological models and try to
interpret the results.

\subsection{Misner-Sharp mass}

Take a collapsing ideal fluid within a compact spherically symmetric
spacetime region described by the following metric in the comoving
coordinates $(t,r,\theta,\varphi)$:
\begin{equation}
ds^{2}=-e^{2\nu(t,r)}dt^{2}+e^{2\psi(t,r)}dr^{2}+R(t,r)^{2}d\Omega^{2}.
\end{equation}
assuming the energy momentum tensor for the perfect fluid in the
form
\begin{equation}
 T^{t}_{t}=-\rho(t,r),~~T^{r}_{r}=p_{r}(t,r),~~T^{\theta}_{\theta}=T^{\varphi}_{\varphi}=p_{\theta}(t,r),
\end{equation}
with the week energy condition
\begin{equation}
 \rho\geq0,~~\rho+p_{r}\geq0,~~\rho+p_{\theta}\geq0,
\end{equation}
we then obtain the Einstein equations in the form
\begin{equation}
 \rho=\frac{2M'}{R^{2}R'}~,~~p_{r}=-\frac{2\dot{M}}{R^{2}\dot{R}},
\end{equation}
\begin{equation}
 \nu'=\frac{2(p_{\theta}-p_{r})}{\rho+p_{r}}\frac{R'}{R}-\frac{p'_{r}}{\rho+p_{r}},
\end{equation}
\begin{equation}
 -2\dot{R}'+R'\frac{\dot{G}}{G}+\dot{R}\frac{H'}{H}=0,
\end{equation}
where
\begin{equation}
G=e^{-2\psi}(R')^{2}~~,~~H=e^{-2\nu}(\dot{R})^{2},
\end{equation}
and $M$ is defined by
\begin{equation}
G-H=1-\frac{2M}{R}.
\end{equation}
The function $M$ can also be written as
\begin{eqnarray} \label{ms-def}
M=\frac{1}{2}\int_{0}^{R}\rho R^{2}dR,
\end{eqnarray}
or
\begin{equation}
M=\frac{1}{8\pi}\int_{0}^{r}\rho
\sqrt{(1+(\frac{dR}{d\tau})^{2}-\frac{2M}{R})}d^{3}V,
\end{equation}
where
\begin{equation}
d^{3}V=4\pi e^{\psi}R'dr,
\end{equation}
and
\begin{equation}
\frac{d}{d\tau}=e^{\nu}\frac{d}{dt}.
\end{equation}
The last form of the function $M$ indicates that when considered as
energy, it includes contribution from the kinetic energy and the
gravitational potential energy. $M$ is called the Misner-Sharp
energy.\\
 Hayward {\cite{hay-ms} showed that in the Newtonian limit of
a perfect fluid, $M$ yields the Newtonian mass to the leading order and
the Newtonian kinetic and potential energy to the next order. In
vacuum, $M$ reduces to the Schwarzschild energy. At null and spatial
infinity, $M$ reduces to the Bondi-Sachs and Arnowitt-Deser-Misner
energies respectively \cite{hay-ms}.

\subsection{Hawking mass}

Hawking {\cite{haw-m} defined a quasi-local mass for the spacelike
topological 2-sphere S:
\begin{eqnarray}
E_{H}(S)=\sqrt{\frac{Area(S)}{16\pi
G^{2}}}(1+\frac{1}{2\pi}\oint_{S}\rho\rho'dS)=\nonumber\\
\hspace{.8cm}\sqrt{\frac{Area(S)}{16\pi
G^{2}}}(\oint(-\Psi_{2}-\sigma\lambda+\Phi_{11}+\Lambda)dS),
\end{eqnarray}
where the spin coefficients $\rho$ and $\rho'$ measure the expansion
of outgoing and ingoing light cones. For the definitions of
$\Psi_{2}$, $\sigma$, $\lambda$, $\Phi_{11}$ and $\Lambda$ see 
{\cite{haw-m}. The Hawking mass has various desirable properties: it
is zero for a sphere in flat spacetime, gives the correct mass for
the Schwarzschild solution on a metric sphere, and tends to the
Bondi mass asymptotically in a static asymptotically flat spacetime.
It is invariant under the boost gauge transformation. It is, however, non-zero 
for generic 2-surfaces in a flat spacetime (see
{\cite{living review} for more detail).

\subsection{Hayward mass}

 Hayward, using a $2+2$ formulation of general relativity, gives  the following definition
 for a quasi-local energy {\cite{hay-m}:
\begin{equation}
 E_{H}(S)=\sqrt{\frac{Area(S)}{16\pi
G^{2}}}[1+\frac{1}{2\pi}\oint_{S}(\rho\rho'-\frac{1}{8}\sigma_{ab}\overline{\sigma}^{ab}-\frac{1}{2}\omega_{a}\omega^{a})dS],
\end{equation}
where $\sigma^{ab}$ and $\overline{\sigma}^{ab}$ are shears and
$\omega_{a}$ is the normal fundamental form. The energy is zero for
any surface in flat spacetime, and reduces to the Hawking mass in
the absence of shear and twist. For asymptotically flat spacetimes,
the energy tends to the Bondi mass at null infinity and to the ADM mass
at spatial infinity. It depends, however, implicitly on the gauge
choice \cite{living review}.

\subsection{Brown-York mass}

Looking into the Hamiltonian formulation of general relativity,
Brown and York {\cite{brown-York} found interesting local quantities
from which the definition of quasilocal mass was extracted. Consider
a 3-dimensional spacelike slice $\Sigma$ bounded by a two-surface
$B$ in a spacetime region that can be decomposed as a product of a
spatial three-surface and a real line-interval representing the
time. The time evolution of the two-surface boundary $B$ is the
timelike three-surface boundary $^{3}B$. When $\Sigma$ is taken to
intersect $^{3}B$ orthogonally, the Brown-York (BY) quasilocal
energy is defined as:
\begin{equation}
E=\frac{1}{8\pi}\oint_{B}d^{2}x\sqrt{\sigma}(k-k_{0}),
\end{equation}
where $\sigma$ is the determinant of the 2-metric on $B$, $k$ is the
trace of the extrinsic curvature of $B$, and $k_{0}$ is a reference
term that is used to normalize the energy with respect to a
reference spacetime, not necessarily flat. This quasi local mass has
the right asymptotic behavior but is not positive in general. The BY
mass depends, however, on the choice of the gauge along the
3-dimensional spacelike slice $\Sigma$. To avoid this gauge dependence , Yau 
{\cite{yau03} (see also {\cite{booth}and {\cite{epp}) introduced a
mass which is gauge independent and always positive. Based on
Yau's definition, Liu-Yau defined the mass
\begin{equation}
E=-\frac{1}{8\pi}\oint_{B}d^{2}x\sqrt{\sigma}(\sqrt{k^{2}-\ell^{2}}-k_{0}),
\end{equation}
where $l$ and $k$ are traces of extrinsic curvatures
$l_{ab}=\sigma_a^c \sigma_b^d \nabla_c u_d$ and $k_{ab}=\sigma_a^c
\sigma_b^d \nabla_c n_d$ respectively, for $\sigma _{ab}=g_{ab}+u_a
u_b-n_a n_b$ being the metric on  the 2-sphere B. It was pointed out 
{\cite{murchadha-prl} that this Liu-Yau mass is strictly positive,
even when the surface is in a flat spacetime. \\
It can be shown that for static spherically symmetric spacetimes the
Brown-York quasilocal energy at the singularity is zero 
{\cite{York06}, in contrast to to the Newtonian gravity, in which
the energy of the gravitational field diverges at the center for a
point particle. Apparently, the nonlinearity of general relativity
has removed this infinity. This QLE attains its maximum inside the
horizon having an infinite derivative just on the horizon, before
matching to its value outside the horizon: the black hole looks like
an extended object.

\section{LTB metric }

 The LTB metric may be written in synchronous coordinates as
\begin{equation}
 ds^{2}=dt^{2}-\frac{R'^{2}}{1+f(r)}dr^{2}-R(t,r)^{2}d\Omega^{2}.
\end{equation}
It represents a pressure-less perfect fluid satisfying
\begin{equation}
\rho(r,t)=\frac{2M'(r)}{ R^{2}
R'},\hspace{.8cm}\dot{R}^{2}=f+\frac{2M}{R}.
\end{equation}
Here dot and prime denote partial derivatives with respect to the
parameters $t$ and $r$, respectively. The angular distance $R$,
depending on the value of $f$, is given by
\begin{eqnarray}\label{ltbe1}
R=-\frac{M}{f}(1-\cos(\eta(r,t))),\nonumber\\
\hspace{.8cm}\eta-\sin(\eta)=\frac{(-f)^{3/2}}{M}(t-t_{n}(r)),
\end{eqnarray}
for $f < 0$, and
\begin{equation}\label{ltbe2}
R=(\frac{9}{2}M)^{\frac{1}{3}}(t-t_{n})^{\frac{2}{3}},
\end{equation}
 for $f = 0$, and
\begin{eqnarray} \label{ltbe3}
R=\frac{M}{f}(\cosh(\eta(r,t))-1),\nonumber\\
\hspace{.8cm}\sinh(\eta)-\eta=\frac{f^{3/2}}{M}(t-t_{n}(r)),
\end{eqnarray}
for $f > 0$.\\
The metric is covariant under the rescaling $r\rightarrow\tilde{r}(r)$.
Therefore, one can fix one of the three free parameters of the metric, i.e.
$t_{n}(r)$, $f(r)$, and $M(r)$.\\
This metric has two generic singularities: the shell focusing
singularity at $R(t,r)=0$, and the shell crossing one at
$R'(t,r)=0$. However, if $\frac{M'}{R^{2}R'}$ and $\frac{M}{R^{3}}$
are finite at $R=0$ then there is no shell focusing singularity.
Similarly, if $\frac{M'}{R'}$ is finite at $R'=0$ then there is no
shell crossing singularity. To get rid of the complexity of the
shell focusing singularity, corresponding to a non-simultaneous big
bang singularity, we may assume $t_{n}(r)=0$, which will be the case
for our toy models. This will enable us to concentrate on the
behaviour of the collapse of an overdensity region in an expanding
universe without interfering with the
complexity of the inherent bang singularity of the metric {\cite{man}.\\
It is easy to show that $\theta_{(\ell)}|_{R=2M}=0$. Therefore, there
may exist an apparent horizon, defined by $R = 2M$, being obviously
a \emph{marginally trapped tube}. It will turn out that this
apparent horizon is not always spacelike and can have a complicated
behaviour for different $r$ {\cite{man}.

\subsection{Misner-Sharp mass}
It is easily seen from (\ref{ms-def}) that $M(r)$ in the LTB metric
is identical to the Misner-Sharp mass. The rate of change of this
mass for any $R=const$ in the collapsing region is positive as can
be seen by the following argumentation. Noting that $\dot{R}<0$ in
the collapsing region, and assuming no shell crossing, $R'>0$, we
obtain from $R'dr+\dot{R}dt=0$ that $\frac{dr}{dt}|_{R=const}>0$.
Therefore, given $\frac{dM(r)}{dr}>0$, we see that
$\frac{dM(r)}{dt}|_{R=const}=\frac{dM(r)}{dr}\frac{dr}{dt}|_{R=const}>0$.

\subsection{Hawking and Hayward masses}
The LTB null tetrad needed to calculate the Hawking mass is given by
\begin{equation}
 \ell^{\mu}=(1,\frac{\sqrt{1+f}}{R'},0,0),~ n^{\mu}=(\frac{1}{2},-\frac{\sqrt{1+f}}{R'},0,0),
\end{equation}
and
\begin{equation}
 m^{\mu}=\frac{1}{R\sqrt{2}}(0,0,1,\frac{i}{sin\theta}),~\bar{m}^{\mu}=\frac{1}{R\sqrt{2}}(0,0,1,\frac{-i}{sin\theta}).
\end{equation}
We then obtain for the Hawking mass
\begin{equation}
 M_{Haw}=M(r),
\end{equation}
i.e. it is equivalent to the Misner-Sharp mass, as expected. This is
due to the vanishing of twist and shear  in metrics being
spherically symmetric for round 2-sphere. We also conclude that the
Hayward mass is identical to Hawking mass for the LTB metrics.

\subsection{Brown-York mass}

The 2-boundary $B$ maybe specified by $r=constant$ and
$t=constant$. We then obtain for the trace $k$ of the extrinsic
curvature $k_{ab}$ for LTB's metric $k=-\frac{2\sqrt{1+f}}{R}$. The
Brown-York energy is then given by
\begin{equation}
M_{BY}=-R \sqrt{1+f}-Subtraction~term.
\end{equation}
The subtraction term is chosen to be the corresponding FRW term for the $t =
constant$ slice, i.e.  $-R \sqrt{1+f}|_{FRW}$.
Now the rate of change of the Brown-York mass is given by
\begin{equation}
\frac{dM_{BY}}{dt}|_{r=const}=-\dot{R} \sqrt{1+f}+(\dot{R}
\sqrt{1+f})|_{FRW}.
\end{equation}
The first term is responsible for the flow of dust falling into the
center and the second term is  due to the cosmological expansion. In
the collapsing phase of the central region the first term is
dominant and the Brown-York mass increases within the sphere of
radius $r$. \\
The Brown-York mass may also be calculated for the 2-boundary $B$
with the constant physical radius $R=constant$ at $t=constant$,
leading to
\begin{equation}
M_{BY}|_{R=const}=-R \sqrt{1-\frac{2M}{R}}-Subtraction~term.
\end{equation}
The subtraction term is again the corresponding FRW term as the
background. \\
Similarly, the Liu-Yau mass for the LTB black hole
model is given by
\begin{equation}
M_{LY}=-R \sqrt{1-\frac{2M}{R}}-Subtraction~term,
\end{equation}
which is valid for $k>\ell$, or for the untrapped region $R>2m$ and
Subtraction~term=$-R$ . The rate of change of this mass is given by
\begin{equation}
\frac{dM_{LY}}{dt}=-\dot{R}
\sqrt{1-\frac{2M}{R}}-\frac{\dot{R}M}{R\sqrt{1-\frac{2M}{R}}}-\frac{d
(Subtraction~term)}{dt},
\end{equation}
which is again an increasing function within the collapsing region.

\section{Mass of evolving black holes within FRW Universe}

We are now interested in the mass of a cosmological overdensity region
evolving into a black hole in a FRW background. We first choose two
toy models and look for the mass of structures they represent. This should
give us an overall view of the different mass definitions and their differences.
Then we go to a more realistic model starting with a given density profile
and look for the model parameters and the corresponding masses.  Our cosmological
black hole is going to be modeled by a LTB solution representing a
collapsing overdensity region at the center and a flat FRW far from
the overdensity region {\cite{man}. The overdensity region may take
part in the expansion of the universe at early times but gradually
the expansion is reversed and the collapsing phase starts. For a more realistic
model we assume the familiar NFW profile {\cite{NFW96} for the
overdensity region within a LTB model and look for its consequences
as regards different mass definitions.

\subsection{Example I:Toy model I with $\lim_{r\rightarrow\infty}f(r)\rightarrow 0^{-}$; structure within an asymptotically closed-flat LTB metric }

The model is defined by the requirement $f(r)<0$ and
$f(r)\rightarrow 0$ when $r\rightarrow\infty$, and $M(0)=0$. In both LTB toy
models we assume $t_b = 0$. Let us
use the ansatz $f(r)=-re^{-r}$ leading to
$$M(r)=\frac{1}{a}r^{3/2}(1+r^{3/2}),$$ where $a$ is a constant having
the dimension $[a]=[L]^{-2}$ {\cite{man}. The constant $a$ is fixed
by $at_{0}=3 \pi /2$, corresponding to the collapsing mass
condensation around $r =0$ starting in the expanding phase of the
bound LTB model. Equations (\ref{ltbe1}) and (\ref{ltbe2}) then lead
to
\begin{eqnarray}\label{met2-1}
R=\frac{\sqrt{r}(1+r^{3/2})}{a e^{-r}}(1-\cos \eta(r,t)),\nonumber\\
\hspace{.8cm}\eta-sin(\eta)=\frac{e^{-\frac{3}{2}r}}{(1+r^{3/2})}at.
\end{eqnarray}
Fig.(\ref{BYmass}) shows schematically the behavior of the curvature
function $f(r)$ and the corresponding Brown-York mass for a sphere
of constant co-moving radius $r = constant$.

\subsection{Example II: Toy model II with $\lim_{r\rightarrow\infty}f(r)\rightarrow 0^{+}$; structure within an asymptotically open-flat LTB metric }

What would happen if we choose the curvature function $f(r)$ such
that it is negative for small $r$ but tends to zero for large $r$
while it is positive? We still have a model which tends to a flat
FRW at large distances from the center, having a density less than
the critical one corresponding to an open FRW model. The model is
defined by the ansatz $f(r)=-r(e^{-r}-\frac{1}{r^{n}+c})$ with $n =
2$ and $c=20000$, leading to {\cite{man}
$$M(r)=\frac{1}{a}r^{3/2}(1+r^{3/2}),$$
where $a$ is a constant having the dimension $[a]=[L]^{-2}$ and
fixed by the requirement $at_{0}=3 \pi /2$. Equations (\ref{ltbe1})
and (\ref{ltbe2}) then lead to

\begin{eqnarray}\label{met2-2}
R=\frac{\sqrt{r}(1+r^{3/2})}{a (e^{-r}-\frac{1}{r^{2}+20000})}(1-\cos\eta(r,t)),\nonumber\\
\hspace{.8cm}\eta-sin\eta=\frac{(e^{-r}-\frac{1}{r^{2}+20000})^{1.5}}{(1+r^{3/2})}at,
\end{eqnarray}
for $f < 0$ and
\begin{eqnarray}\label{met2-2}
R=\frac{\sqrt{r}(1+r^{3/2})}{a (\frac{1}{r^{2}+20000}-e^{-r})}(\cosh \eta(r,t)-1),\nonumber\\
\hspace{.8cm}\eta-sinh\eta=\frac{(\frac{1}{r^{2}+20000}-e^{-r})^{1.5}}{(1+r^{3/2})}at,
\end{eqnarray}
 for $f > 0$.

Fig.(\ref{BYmass}) shows schematically the two Brown-York masses
inside spheres of constant comoving radius $r$ for the closed-flat
and open-flat cosmological black hole toy models. Note the negative
values of the BY mass in the case of open-flat model at distances
far from the central overdensity region. In the case of closed-flat
model, the Brown-York mass behaves similar to the corresponding
Schwarzschild mass {\cite{York06}. In both cases the Brown-York mass
is zero at the central singularity in contrast to the Misner-Sharp
and Hawking and Hayward mass, and remains finite within and on the
horizon. \\
\begin{figure}[h]
\begin{center}
\includegraphics[width = \columnwidth]{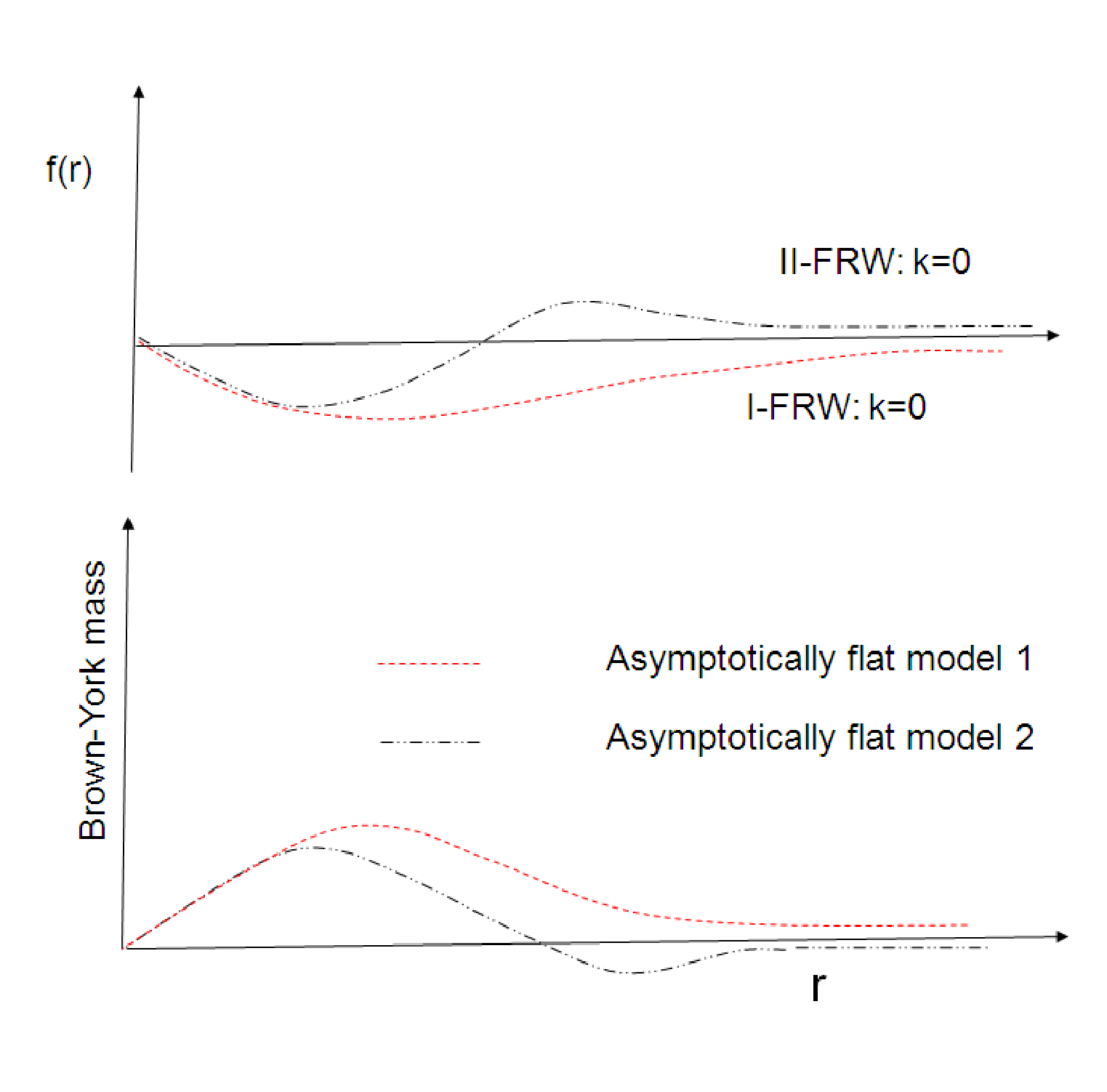}
\caption{ \label{BYmass} The upper diagram shows curvature functions
$f(r)$ for the closed-flat (I) and open-flat models (II). The lower
diagram shows schematically two Brown-York masses for constant
comoving radius $r$.}
\end{center}
\end{figure}

The more interesting case of the mass inside a sphere of constant
physical radius $R$ is shown in Fig.(\ref{BYmass2}) just for the
closed-flat case. The corresponding mass behaviour for the open-flat
case is similar to the closed-flat one, at least in the range of
radius plotted in Fig.(\ref{BYmass2}). Note the $\Omega-$value
($\Omega= \rho/ \rho_{cr} $) as a function of $R$ plotted in the
same figure. A comparison to the BY mass of the Schwarzschild metric 
{\cite{York06} shows following similarities and differences. The
Brown-York mass of our cosmological black hole toy model has two
maxima: one near the apparent horizon and the other far from the
horizon at cosmological distances, corresponding to the FRW
cosmological horizon, and at the same time around a void occurring
in the models studied. This void is, however, too shallow to be seen
in the figure. We will see in the next section that in more
realistic cases this second maximum splits in two different ones:
one around the void where the density tends to the background value
$\Omega = 1$, and the other around the cosmological horizon which is
out of the range of our interest. The first maximum occurs just
before the apparent horizon of the central black hole followed by an
infinite slope similar to the BY mass of the Schwarzschild metric 
{\cite{York06}. Let us call it the \emph{Horizon mass maximum},
which can be seen to be related to the non-zero Misner-Sharp mass of
the black hole. After reaching a minimum, the BY mass then increases
again with increasing physical radius $R$ up to the second maximum,
in contrast to the BY mass for the Schwarzschild case which
decreases to the ADM mass. This second maximum, however, occurs
after an infinite slope just after the cosmological particle
horizon, in contrast to the behavior of the horizon mass maximum
which has the infinite slope before the mass maximum and the
apparent horizon. Because of the fact that this maximum occurs
around the region where the density have reached the background FRW
density, after passing a void and separating the central black hole
from the almost homogeneous background, we call it the
\emph{structure mass maximum}. In contrast to the one corresponding
to cosmological distances which we may call \emph{cosmological mass
maximum}. the separation of these two maxima will be obvious in the
example III. Note that The Liu-Yau mass behaves similar to the
Brown-York mass in the strong gravity region, but it approaches the
Misner-Sharp mass far
from the center.\\
Fig.(\ref{BYmass4}) shows the BY mass at three different times, just
before the onset of singularity and after the singularity has
appeared, corresponding to different density profiles. It is obvious
from the figure that the mass maxima increase and shift towards
larger $R$ values as the time increases. The Horizon maximum is
missing before the onset of the singularity.  This maximum may be
used as an indicator of singularity appearance in numerical
relativity. The structure mass maximum is, however, present for any
density profile irrespective of the occurrence of a central
singularity (see Fig.\ref{BYmass3}).
\\
\begin{figure}[h]
\includegraphics[width = \columnwidth]{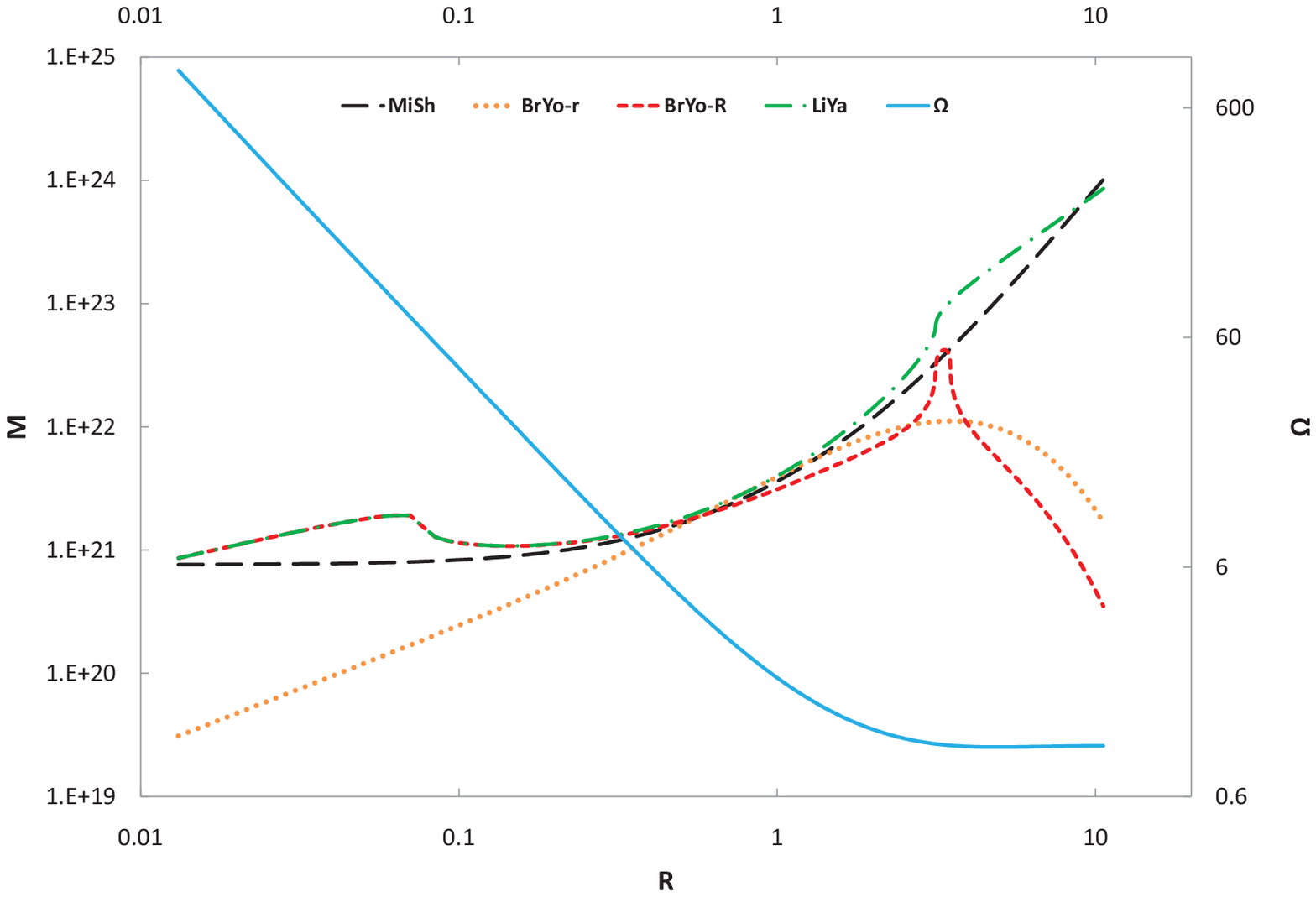}
\caption{ \label{BYmass2} Different masses of the cosmological
black hole toy model I for constant physical radius $R$. Horizon radius is about $0.1$ in terms of $R$ units.}
\end{figure}
\\
\begin{figure}[h]
\includegraphics[width = \columnwidth]{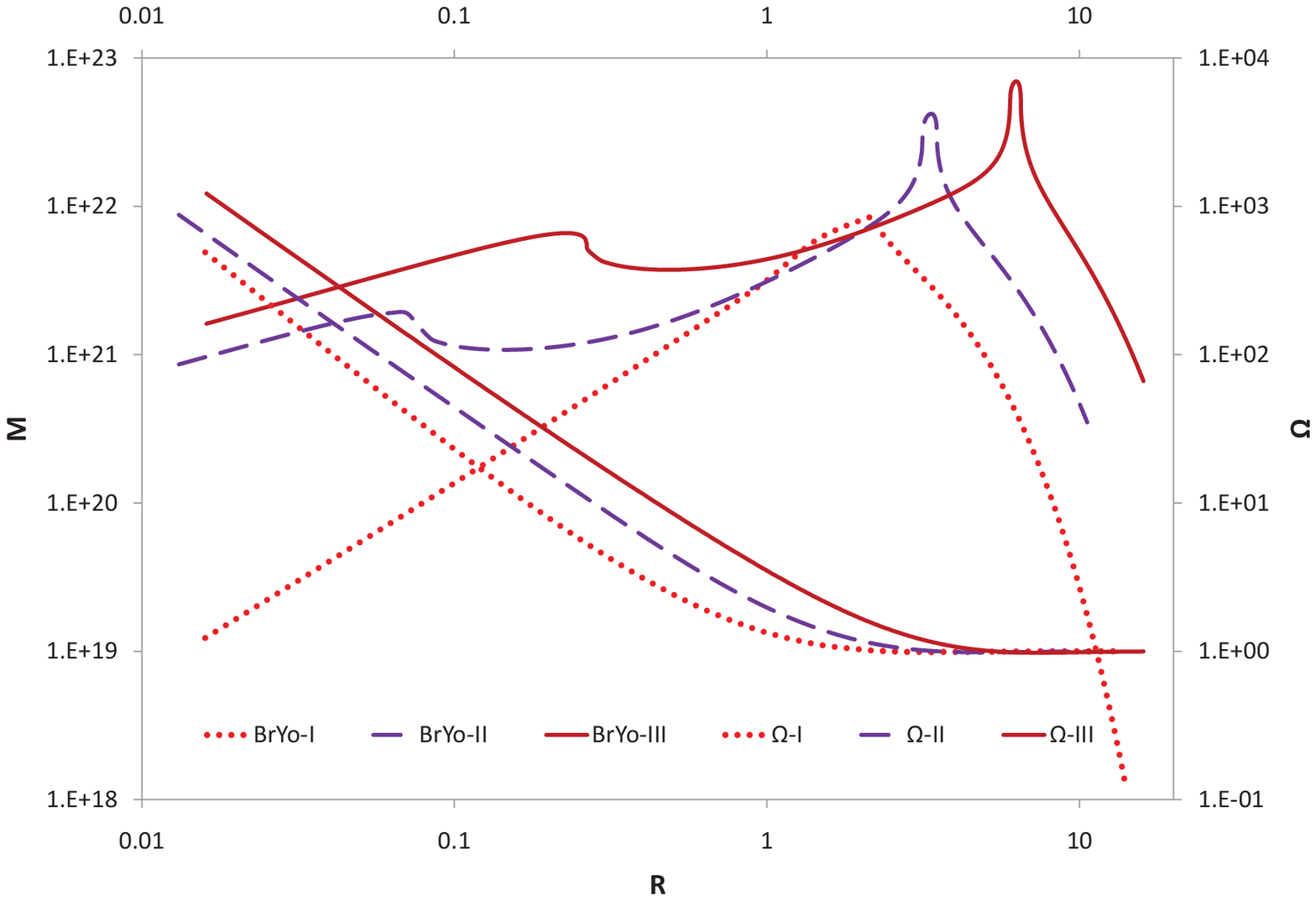}
\caption{ \label{BYmass4} Brown-York mass for three density profiles
corresponding to three different times after the onset of the
singularity and a time before the singularity has been formed. As
the mass in-fall increases the black hole mass maximum increases too
and shifts towards larger $R$ values. For dashed line, horizon radius is about $0.1$ in terms of $R$ units.}
\end{figure}

\subsection{Example III: Structures within asymptotically flat LTB models
having a NFW density profile}

For a more realistic modeling of a cosmological structure, we need
an algorithm to construct functions $M$, $f$, and $t_b$ in the LTB
solution from physical quantities of the system such as the density
profile. Krasi\'{n}sky and Hellaby {\cite{KrHe} propose an algorithm
by which knowing the initial and final density profiles of an object
one can find $f$ and $t_b$ as functions of $M$. For the sake of simplicity, it
has been assumes here $M = r$.   \\
We choose a Gaussian profile for the density at the last scattering surface
as the initial time. The final profile, say at $z \sim 0.2$, is then chosen to
be the universal density profile for the dark matter suggested by Navarro, Frenk
and Wright {\cite{NFW96}. To simulate a void compensating the overdensity mass
region relative to the cosmological background \cite{KhMan}\cite{MatHum}, we convolute density profile of
structure with a Gaussian underdensity. At far distances from the center of
structure, density tends to the critical density, corresponding to a flat matter
dominated cosmological background:
\bqn
\rho_{i}(r)=&&\rho_{crit}(t_{i})((\delta_{CMB}e^{-(\frac{r}{R_{i1}})^2}
-b_1)e^{-(\frac{r}{R_{i2}})^2}+1)\nonumber\\
\rho_{NFW}(r)=&&\rho_{crit}\frac{\delta_c}{(\frac{r}{r_s})(1+\frac{r}{r_s})^2}\nonumber\\
\rho_{f}(r)=&&\rho_{crit}(t_{f})((\frac{\delta_c}
{(\frac{r}{r_s})(1+\frac{r}{r_s})^2}-b_2)e^{-(\frac{r}{R_{f}})^2}+1)\nonumber
\eqn
Using this algorithm, we have calculated the corresponding LTB
functions, needed to define the metric and to calculate different
masses. $f$ as a function of $M$ is depicted in Fig.(\ref{fm}). Note that, in contrast
to toy models discussed above, now the bang time $t_b$ is non-vanishing as is shown Fig.(\ref{tbm}). Theses
functions have been calculated for $r$-values larger than a minimum corresponding to the central
singularity $R = 0$. To calculate the mass we have assumed $r_s \approx 500 kpc$ and $\delta_c \approx 4,000$ {\cite{wrbr}.
\begin{figure}[h]
\includegraphics[width = \columnwidth]{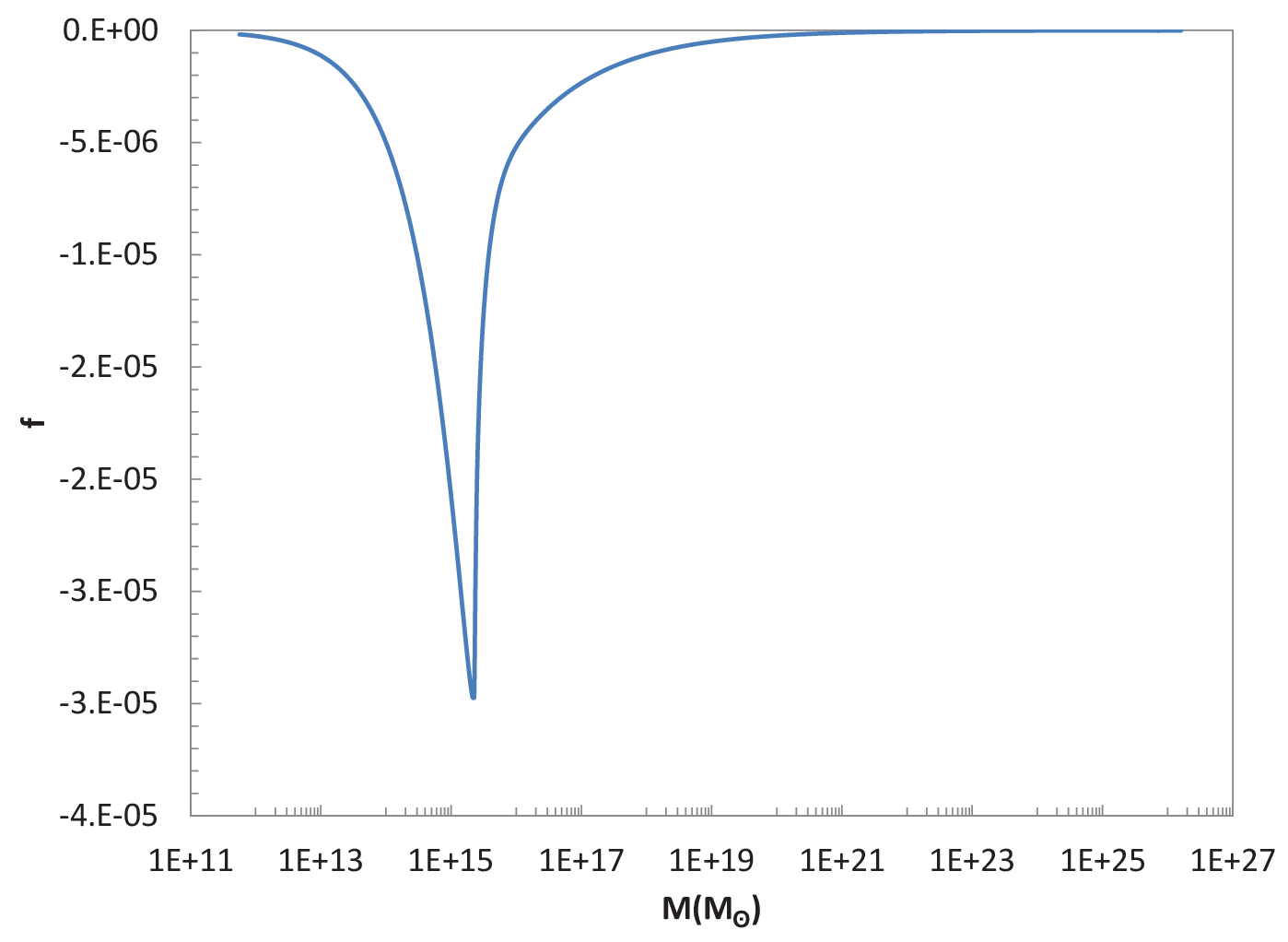}
\caption{ \label{fm} $f(M)$ for a cluster of galaxies with NFW density
profile in a flat background. }
\end{figure}
\\
\begin{figure}[h]
\includegraphics[width = \columnwidth]{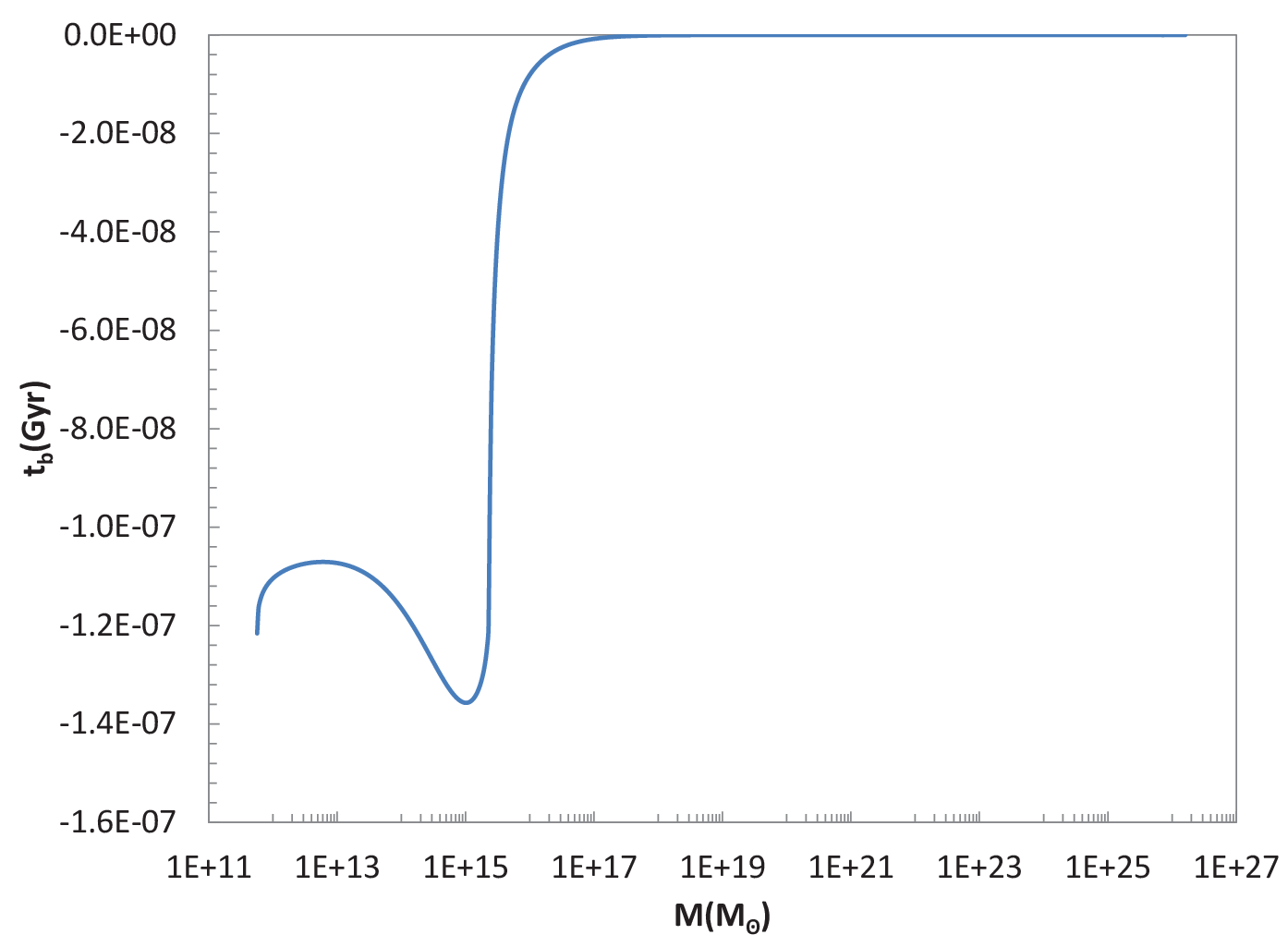}
\caption{ \label{tbm} $t_b(M)$ for a cluster of galaxies with NFW density
profile in a flat background.}
\end{figure}

The result for different mass definitions is depicted in
Fig.(\ref{Masses-cl}) as a function of the physical distance from
the center. The mass up to the void turns out to be of the order of
$10^{14}$ solar masses, which turns out interestingly to be
of the same order for all mass definitions. \\

There are some interesting features in the Fig.(\ref{Masses-cl}).
Note first that the horizon mass maximum, occurring also for the NFW
profile, is not depicted in the Fig.(\ref{Masses-cl}) as it is very
near to the origin of the figure. Numerical calculation shows that
the apparent horizon is at a physical distance of the order of
$10^{11} km = 0.01 kpc$ corresponding to a Misner-Sharp mass of the order of
$3\times10^{11} M_{\odot}$. The structure maximum is shifted more to the left
at distances of the order of few Mpc corresponding to the place of the void just
before the density profile reaches the background values. Different
mass definitions seems to coincide at this structure maximum which
is about $10^{14}$ for the model we have constructed and it is about
$300$ times the horizon maximum mass and about 30 times
the mass up to the distance of about 1Mpc, corresponding to $\Omega \approx 100$.\\
The \emph{cosmological mass maximum} of Brown-York, which appears at
cosmological distances, is now separated from the structure mass
maximum. It is interesting to note that BY mass for constant
co-moving and physical radius remains almost the same for distances
after the void, while the Misner-Sharp and Liu-Yau masses increase,
being almost equal. Fig.(\ref{Masses-gal}) shows the corresponding diagram
for galactic masses. We recognize similar features as those for
cluster masses, except the less exposed void at distances less
than one Mpc. \\
\begin{figure}[h]
\includegraphics[width = \columnwidth]{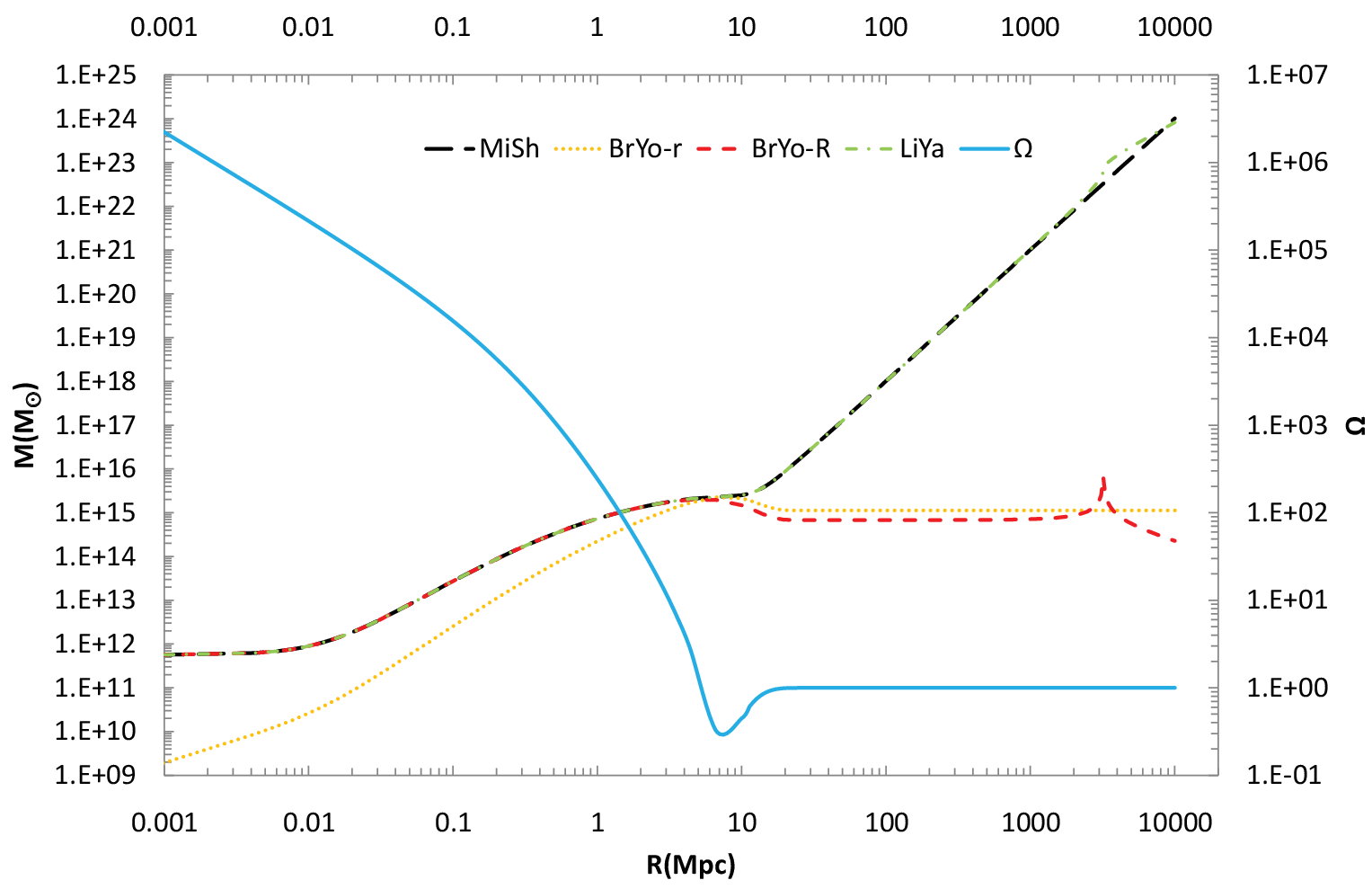}
\hspace*{10mm}\caption{ \label{Masses-cl} Misner-Sharp, Brown-York,
and Liu-Yau masses for a cluster of galaxies with NFW density
profile in a flat background. The total density parameter $\Omega=
\rho/ \rho_{cr} $ as a function of the physical radius is also
shown. Horizon radius is about $0.01pc$.}.
\end{figure}
\begin{figure}[h]
\includegraphics[width = \columnwidth]{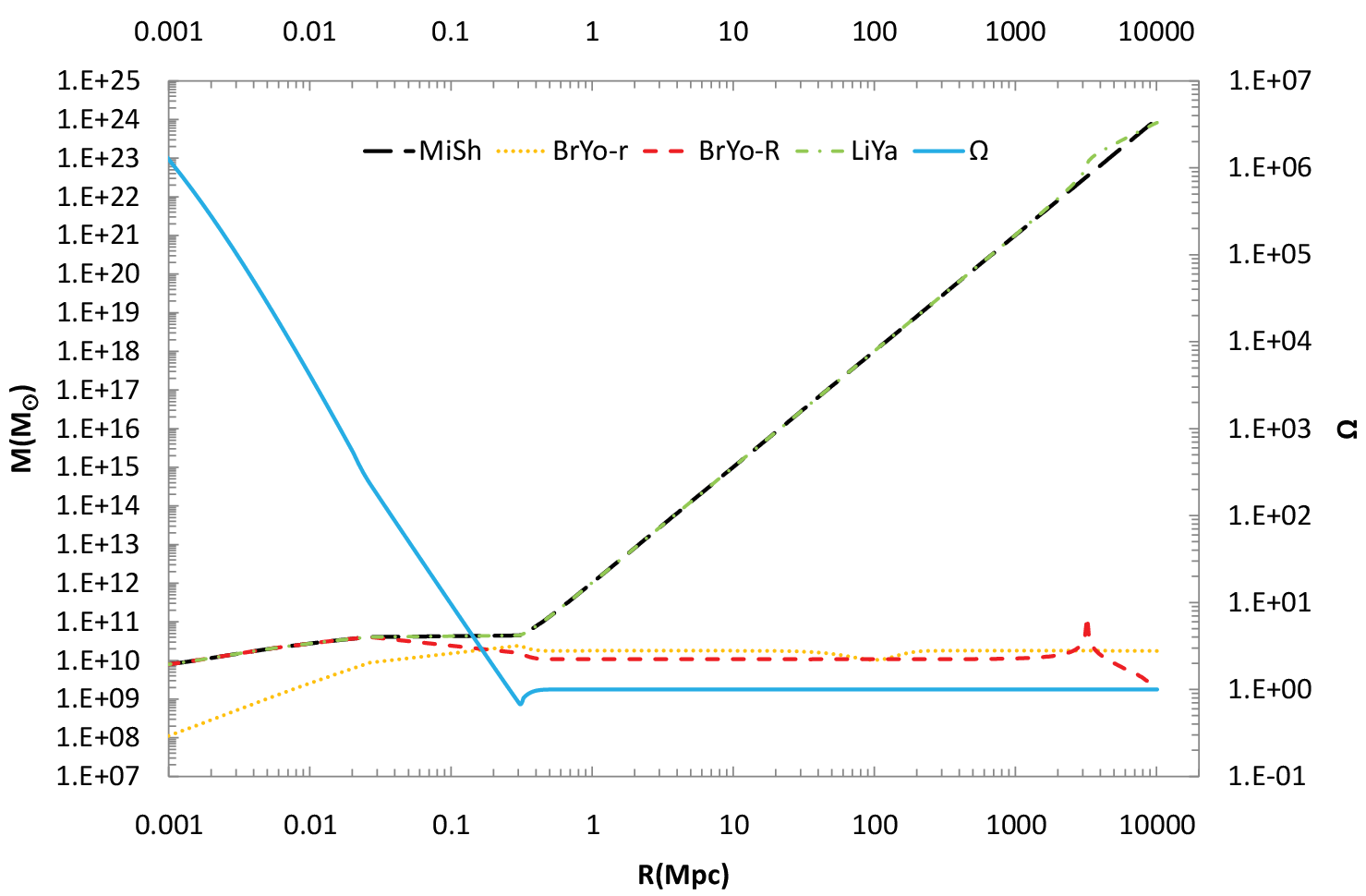}
\hspace*{10mm}\caption{ \label{Masses-gal} Misner-Sharp, Brown-York
and Liu-Yau masses for a galaxy with NFW density profile in a flat
background. The density parameter $\Omega= \rho/ \rho_{cr} $ is also
shown. Horizon radius is about $10^{-4}pc$.}
\end{figure}

The rate of matter flux through the apparent horizon (as a
quasi-local black hole boundary in an asymptotically flat universe)
is a useful quantity in astrophysical studies of black holes as well
as theoretical study of black hole laws \cite{edingtion-limit}. The
matter flux for Misner-Sharp mass along the apparent horizon is
given by
$\frac{dM(r)}{dt}|_{AH}=\frac{dM(r)}{dr}\frac{dr}{dt}_{AH}$. As we
see in Fig.(\ref{dmdt}), the rate of the matter flux increases with
time at the initial phase of the black hole formation, up to a
maximum value of the order of magnitude $10^6$ solar masses. It then
decreases while the collapse is continuing and the black hole
boundary (apparent horizon) freezes out in the expanding background
(see Fig.\ref{Horizon}). In the case of BY mass, the rate of matter
flux is similar to that of Misner-Sharp one (see Fig.(\ref{dmdt})),
being almost twice as much on the horizon as in the former case.
\begin{figure}[h]
\includegraphics[width = \columnwidth]{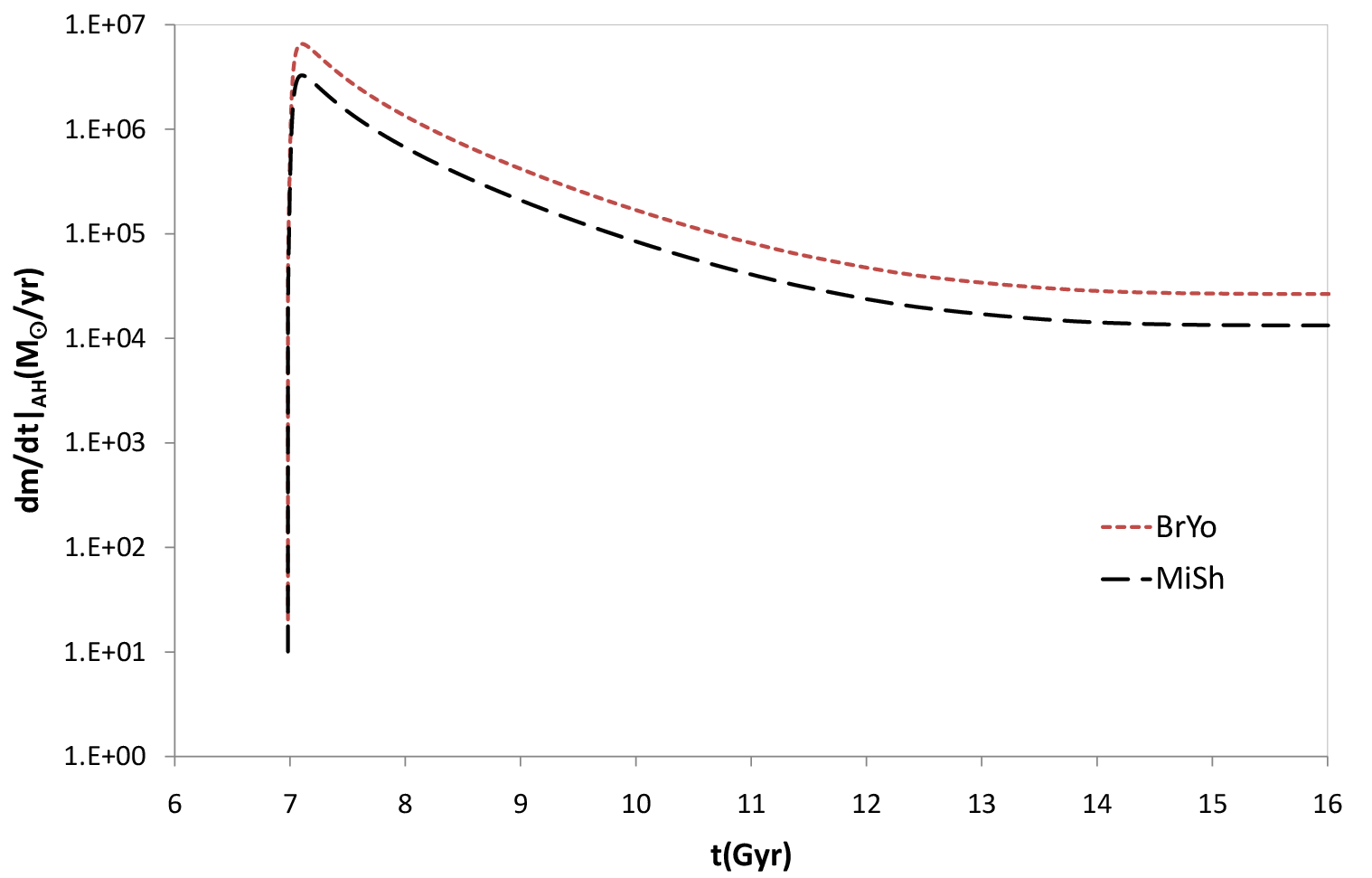}
\hspace*{10mm}\caption{ \label{dmdt} The Misner-Sharp and Brown-York
matter flux computed along apparent horizon. Note the rapid increase of
the matter flux  rate up to the 'freezing point' point of the Horizon
and the following decrease.}
\end{figure}
\\
\begin{figure}[h]
\includegraphics[width = \columnwidth]{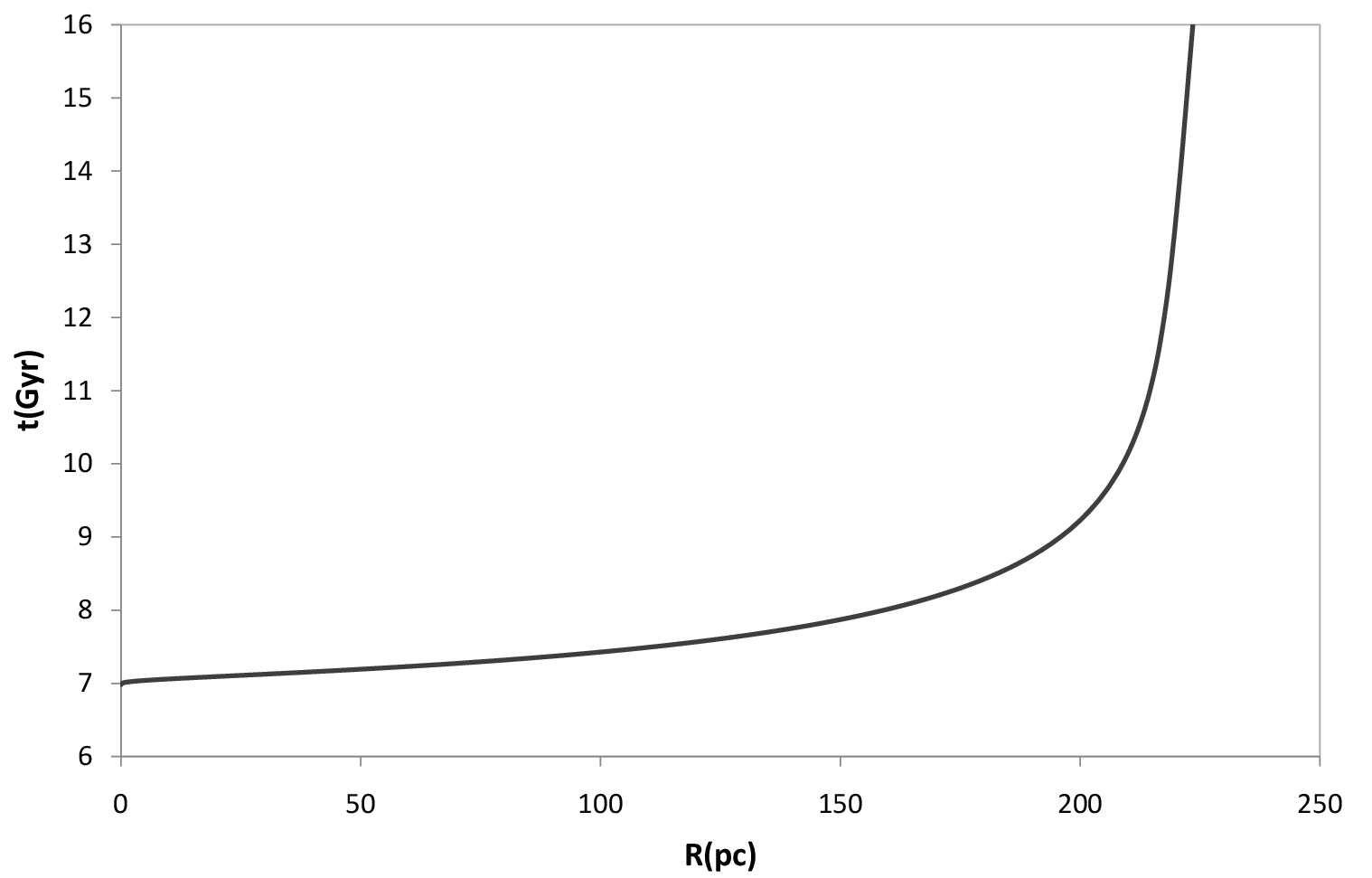}
\hspace*{10mm}\caption{ \label{Horizon} The apparent horizon line in
t-R plane.}
\end{figure}

 Fig.(\ref{BYmass3}) compares the BY mass
for two different NFW density profiles at a given time after the
onset of the singularity. It shows the increase of the horizon and
black hole or structure masses, and their shift to the right with
the increasing density profile.
\begin{figure}[h]
\includegraphics[width = \columnwidth]{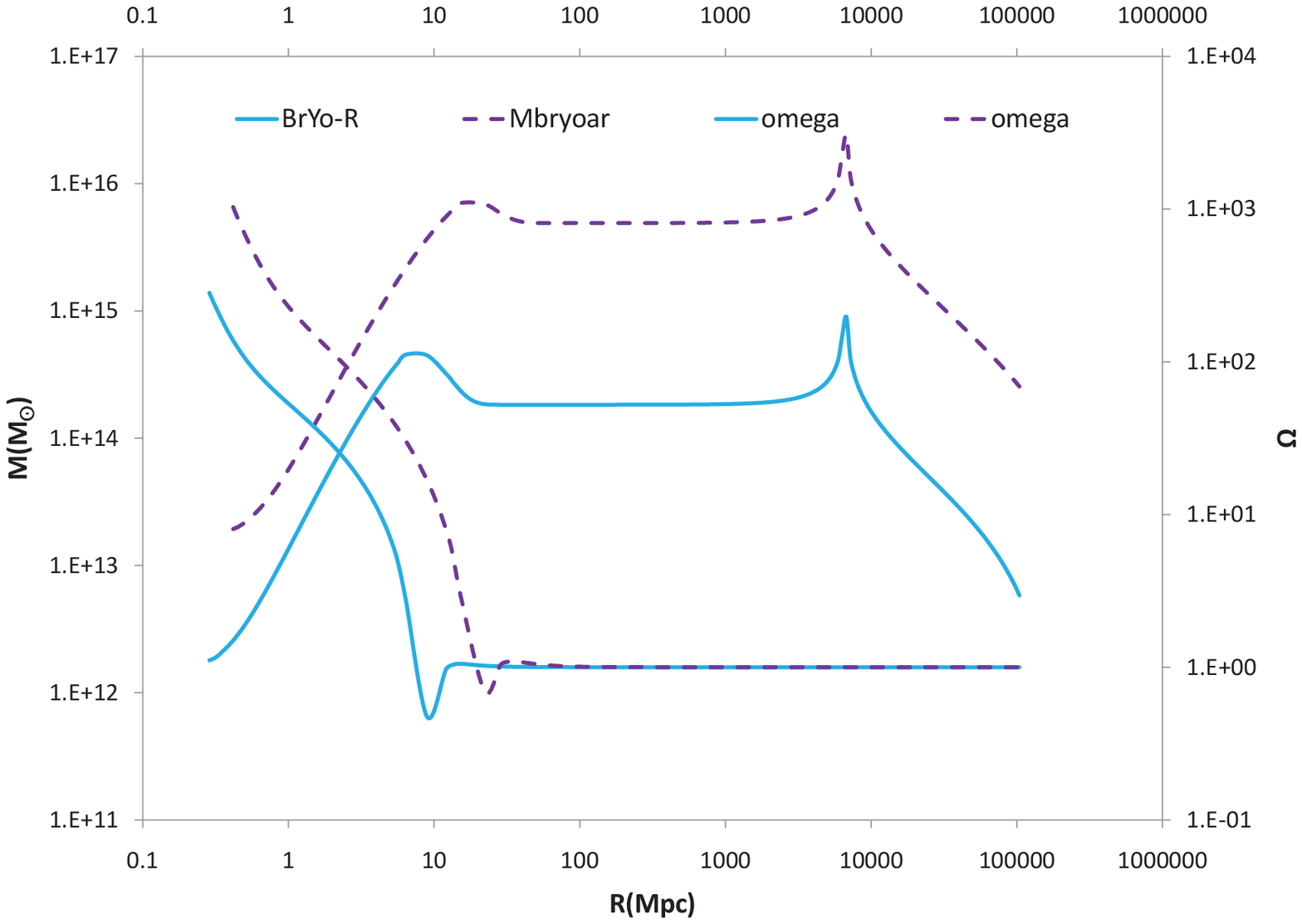}
\caption{ \label{BYmass3}Behavior of Brown-York mass for two
different density profiles at the same time.}
\end{figure}

\section{Discussion and conclusions}

Being faced with the challenge to define mass of structures in a
dynamical setting within general relativity, we have used models of
mass condensation within a dynamical cosmological background to gain
concrete insights of the similarities and differences between some
of the more familiar mass definitions. The first two toy models
representing collapsed overdense regions within asymptotically
open-flat and closed-flat FRW models, show similar behavior at
distances not very far from the collapsed region. The Brown-York
mass, however, becomes negative for the open-flat model
(Fig.\ref{BYmass}) at large distances relative to the place of the
void where the density reaches the background one. For both models
the BY mass tends to zero at infinity, as expected. It seems that
independent of any density profile as the initial condition, if we
wait enough, there is always a void before the density reaches the
background value. At about the same distance, there is always a
maximum of the mass, the 'structure mass maximum', which increases and
move to the larger distances from the center as the density profile
increases with the time through more infall of matter to the
singularity. At the central singularity for $R = 0$ the Brown-York
mass is zero in accordance with the BY mass of the Schwarzschild
metric but in contrast to the Misner-Sharp, Hawking, and Hayward
mass. At distances less than the void the BY mass is less than the
Misner-Sharp mass. It is interesting to note that all mass
definitions lead to almost the same value near the structure maximum.\\
In the case of the more realistic NFW density profile these
features are even more distinguished. As can be seen from the
Fig.(\ref{Masses-cl}), the structure maximum moves to the left at
distances of the order of few Mpc for masses of the order of clusters.
This is the same physical distance at which all three masses almost coincide
and are equal to each other. At distances above $10Mpc$ the LY and MS mass are almost
equal except for a maximum LY mass at distances corresponding to the
cosmological BY maximum mass and differing from the Misner-Sharp
ones. The BY mass at distances larger than place of the void may be
defined as a function of the comoving radius or physical radius,
being almost the same. It differs, however, substantially from the
LY and Misner-Sharp mass, and remains almost constant up to large
cosmological distances before reaching the last cosmological
maximum. \\

The results obtained so far is indicative enough that the mass definitions may differ
substantially. It is not said, however, that it may have any impact on our astrophysical
mass determinations. In fact the mass definition in astrophysics is not as trivial as
it may seems in the Newtonian dynamics. It is not even clear that there is any need at all
for the concept of 'mass' in any astrophysical or cosmological setting. We are currently using
our models to see if one can see any discrepancy between the general relativistic mass
definitions and the one in Newtonian dynamics. Given the the Newtonian approximation for
weak fields in general relativity, it is a legitimate question if this limit may also be used at
cosmological distances {\cite{GrWa}. We have already seen how
different Misner-Sharp mass is relative to BY one! To tackle such questions we are currently
applying our model to gravitational lensing phenomena {\cite{ltb-lens} and the rotation curves
of point masses within dynamical structures to see any deviations from the Newtonian approximation.
Specifically, we model a lens as a structure in the cosmological background and solve the geodesic
equations numerically in a general relativistic framework using our model structure. Note that
the structure maximum mass occurs at points where the density is of
the order of $10^{-29} g/cm^3$ and the gravity is weak enough to assume
the Newtonian approximation. It is, however, not clear that we can ignore
the nonlinear effects of general relativity at such large distances \cite{GrWa}.

\end{document}